# Dynamic Verification for File Safety of Multithreaded Programs

**Mohamed A. El-Zawawy†,†† and Nagwan M. Daoud††**,

†College of Computer and Information Sciences, Al-Imam M. I.-S. I. University, Riyadh, Kingdom of Saudi Arabia
††Department of Mathematics, Faculty of Science, Cairo University, Giza 12613, Egypt

**Summary**
In this paper, we present a new semantics to check *file safety* of multithreaded programs. A file-safe program is one that reaches a final configuration under the proposed semantics. We extend the *While* language with file operations and multi-threading commands, and call the new language $while_f$. This paper shows that the file safety is an un-decidable property for $while_f$. The file safety becomes a decidable property in a special case shown in this paper. The case happens when users provide pointer information. If the file is safe we call it a *strongly safe file* program. We modify the syntax and the semantic of the language and called it $SafeWhile_f$.

## 1. Introduction

When working with files it is very important to avoid any unwanted access to it, whether it was to read or write from and into it [11]. File safety is a crucial property for programs. There is also some other things that should be considered when working with files, like making sure to read from an opened file. These things include not to open a file that is already opened, and similarly for closing a file [21]. In this paper we focus on the value returned by the file pointer which gets increased by one automatically every time we read from a file. This paper studies the safety in general and discusses the un-decidability of files safety, even for terminating programs. So we show that if we control the value of file pointer the file safety of terminating programs becomes a decidable property. To control the pointer value, the value will be passed to the read command. We ignore the end of file error. By controlling the pointer value the safety of accessing a file becomes decidable. The work is done in a dynamic style, meaning that present a dynamic way to check file safety.

**Dynamic VS static file safety**.
*Static File Safety*. Static analysis is the type of analysis that is performed on the program before executing it. It is done mostly on source code using any static checker like type systems. Static analysis doesn't need any test cases to find errors. It usually searches for syntax errors or type errors etc. Also it doesn't care what the program should do and if it does the required task. Static analysis has different tools to find different errors like PMD, ESC/Java, and MALPAS. The last one uses the direct graph to specify if the program meets a mathematical specification. Reject a lot of programs for harmless mistakes can be considered as the main disadvantages of this type of analysis.
*Dynamic File Safety*. Dynamic analysis means the opposite of static, so here the program is being checked by the compiler during the run time. The idea of checking the file safety in a dynamic way is to design a semantics that will get stuck in a non-final state to report an error, otherwise it will terminates in a final configuration.

**Motivating example.** Consider the following example:

1. *open (f);*
2. *forkfor{ (x,p) = read (f); y=p}*
3. *close (f);*

At line 1, the file *f* is opened and the pointer value is set to 0. At line 2 the two statements will be executed for a finite number of times. The *forkfor* command takes a statement and repeats it a finite number (dynamically fixed) of times. The two statements are for reading from a file *f*, storing the read value in *x*, and storing the pointer value in *p*. Finally the value of *p* is assigned to the variable *y*. When the fork loop ends the program closes the file at line 3. In this example suppose we are interested in the value of *y* which is increased by one every time the *read(f)* is executed. Because we do not know in advance how many times the *fork* will be executed, the value of *y* is not deterministic. This is the source of un-decidability of file safety problem in multithreading programs. We show that providing the file pointer as a parameter for the read statement turns the problem into a decidable one for some cases.
In this paper, we work with an extension of *while* language. We called the extend language $while_f$ which contains commands for file operations and for parallelism concepts.

As we mentioned before that un-determinism comes from pointer values returned by reading from a file. To control the pointer value and eliminate the un-determinism we modified $while_f$ to get the language $SafeWhile_f$. We say that a program is *strongly file safe* if it terminates at a final configuration under the semantics rules defined for $SafeWhile_f$. Otherwise, the file is not safe.
Our contribution is as follows:

- The paper presents a new language model, $while_f$ with its formal semantics, which is an extension of *while* language.
- We prove that even for terminating programs the file safety remains un-decidable property.
- We modified $while_f$ to get $SafeWhile_f$ and discuss the decidability in this restricted definition.
- We proved that file safety becomes decidable under the definition of $SafeWhile_f$.

The paper is organized as follows. In section 2, we present a formal semantics for $while_f$ with file operation statements (open, read, and close a file). In section 3, we discuss file safety even for terminating programs in $while_f$ and present a formal definition for file safety. In section 4, we introduce $SafeWhile_f$ with a restriction to ensure file safety, and prove our final result. Finally we conclude the paper Section 5.

## 2. Formal Semantics of $while_f$

In this section, we define $while_f$ using rewriting logic semantics [19, 20, 21, 18]. A language $L$ can be defined via the triple ($\sum_L$, $E_L$, $R_L$), where:
$\sum_L$ is the syntax of $L$; $E_L$ is a set of $\sum_L$ equations. These equations have no computational meaning; they only transform the statements to a form in which the rules can be applied on. $R_L$ is the set of $\sum_L$ rules. Every rule have two sides, we can replace the left side with the right one of the same rule in one way direction. We write $R \vdash t=t'$ to express that it is possible to prove using some rules that $t$ and $t'$ are equivalent. The expression $R \vdash t \rightarrow t'$ means that $t'$ can be derived from $t$ using only one rule. And $R \vdash t \rightarrow^* t'$ means that we can reach $t'$ starting from $t$ in many steps. In any derivation we write = to express that we use an equation and → to express that we use a rule. Figure 1 shows a complete definition of $while_f$.
**$while_f$ Syntax**. The syntax as shown in figure 1 has two sets: $A$ which stands for atomics statements and $S$ standing for non-atomic ones. When we say atomic we mean a statement that can't be composed from other statements like integers, variables, binary operations, etc. The set $S$ contains all the atomics and also the statements that can be composed from atoms. For integers $n \in Z$, pointers $p \in N$, variables $x \in Var$, the set of variables names. File names $f$ rang over the set $F$ which contains the entire files name. We let *op* denote any arithmetic operation like +,-,*, /, also can be overloaded to include Boolean operations like <=,>=,<>,==,!=. The syntax also include *if* statement and *while* statement. File operations *open, close,* and *read* are also included. When working with files, say *open* a file $f$ the pointer is set to 0. Reading from a file *read(f)* returns two values; the first is the read value and second is the pointer value which will be increased by one automatically. Finally to close a file say $f$ we use *close* statement with passing to it the file name $f$. The *skip* statement is included within the atomic statements. $S$ covers all the atomics, sequence, and some statements for multithreading concepts context. For multithreading, we include *fork, fork for*, and *fork if*. We note that *fork* takes a number of statements and execute them in parallel. The command *forkfor* takes only one statement and repeats it for a finite number of times also in parallel. The *forkif* do the same job as *fork* but only if the condition is satisfied. The syntax of function call and local variables are also included in our language model.

**Definition 1.**
A computation in $while_f$ is called *well terminating* if and only if it is equal to "." - the unit- or to an integer value $n$.

**Desugaring Equation**. These equations do a very simple task. The equations transform the statement into a form which can be easily found in the left hand side of a rule. It transforms & and || to the *if* statement using ordinary logic rules.
**Configuration**. We follow [21] in designing configurations which consist of 3 entries: $< \dots >_S$ which holds the context as a sequence. The second entry $< \dots >_{env}$ holds the environment which maps variables to integers. And the entry $< \dots >_{fs}$ to represent file status table [11] which is a map from file names to $\{o,c\}$ to decide wheatear the file is opened or closed. We mentioned before that the context was treated as sequence, so we use $\mapsto$ to mean that. The symbol □ denotes a frozen operator. This operator replaces some context in a rule to denote that the original context is being calculated.

**Definition 2**.
Let configurations of the form $<< \dots >_S , < \dots >_{env} , < \dots >_{fs} >$ called *concrete configuration*, and let G be the set of all configurations. We have several types of such configurations:

- Initial configuration which take the form $<< S >_S , < \dots >_{env} , < \dots >_{fs} >$ which also can be written as *[S]*.

- Final configuration which take the form $<< .>_S$ , $<\rho>_{env}$ , $< \sigma >_{fs}>$ or $<< n>_S$ , $<\rho>_{env}$ , $<\sigma >_{fs}>$.
- A normal form configuration is a configuration that can't be rewritten any more, i.e. there is no $\tau'$ such that $while_f \vdash \tau \rightarrow \tau'$.
- Stuck configuration is a configuration in which we reach a normal form but it is not final.
- Terminating configuration in which we proceed for a finite number of times.

**Semantics Rules**. All the rules are easy to read except for some rules that will be illustrated below. For the rule open there is a side condition for the file to be closed. The same idea applies for the *close* rule; the side condition states that the file status should be *o*. So we can't open (close) a file that is already opened (closed). To read from a file the *read* rule has a side condition to ensure that the file is already opened. In the *read* rule we can see that it returns two values, the first is the value read which is stored in *x*. The second value is the pointer value which is stored in *p*. After reading from a file the environment updates both *x* and *p* with the new values. Note that to update the value of *x* we use a function named φ. The definition of this function is $\varphi:\{F,N\}\rightarrow Z$. This function takes a file name and pointer value, and returns the value stored at that place. The sequence rule is straightforward.

***fork* rule**. The fork rule is tricky. Let every $S_i$ , $1\leq i\leq n$ appearing in *fork* consists of some atomic expressions say $a_{ij}$, $j\in N$. We define a permutation θ from N to itself [10], i.e we break each *S* into its atomics. Then we execute all the atoms of all *S*'s in *fork* according to the order returned by the permutation θ.

*forkfor* and *forkif* both are executed using fork semantics. This is so as *forkfor* executes *S* for a finite number of times and *forkif* executes $S_i$ only if $a_i$ is satisfied.

**Syntax:**

$A \ni a ::= n \mid x \mid a_1 \; op \; a_2 \mid a_1 \; \&\& \; a_2 \mid a_1 \mid\mid a_2 \mid a_1 = a_2 \mid$ $if \; a \; then \; a_t \; else \; a_f \mid while \; a \; do \; a_t \mid$
$open\;(f) \mid x = read\;(f,p) \mid close\;(f) \mid skip$
$S ::= a \mid a_1;a_2 \mid fork\;\{S_1,S_2,\ldots, S_n\} \mid forkfor\;\{S\} \mid forkif\{(a_1,S_1),(a_2,S_2),\ldots,(a_n,S_n)\}$

**Desugaring equations**

$a_1 \; \&\& \; a_2 = if \; a_1 \; then \; a_2 \; else \; 0$
$a_1 \mid\mid a_2 = if \; a_1 \; then \; 1 \; else \; a_2$

**Semantics rules**

$<x \mapsto S>_s< \rho >_{env}< \sigma >_{fs} \rightarrow <n \mapsto S>_s< \rho >_{env}< \sigma >_{fs}$ $if \; \rho(x)=n$

$<a_1 \; op \; a_2 \mapsto S>_s< \rho >_{env}< \sigma >_{fs} \rightarrow <a_1 \mapsto \square \; op \; a_2 \mapsto S>_s< \rho >_{env}< \sigma >_{fs}$

$<n \; op \; a_2 \mapsto S>_s< \rho >_{env}< \sigma >_{fs} \rightarrow <a_2 \mapsto n \; op \; \square \mapsto S>_s< \rho >_{env}< \sigma >_{fs}$
$<n_1 \; op \; n_2 \mapsto S>_s< \rho >_{env}< \sigma >_{fs} \rightarrow <n \mapsto S>_s< \rho >_{env}< \sigma >_{fs}$ $if \; n_1 \; op \; n_2=n$

$< a_1 \; \&\& \; a_2 \mapsto S>_s< \rho >_{env}< \sigma >_{fs} \rightarrow < if \; a_1 \; then \; a_2 \; else \; 0 \mapsto S>_s< \rho >_{env}< \sigma >_{fs}$

$< a_1 \mid\mid a_2 \mapsto S>_s< \rho >_{env}< \sigma >_{fs} \rightarrow < if \; a_1 \; then \; 1 \; else \; a_2 \mapsto S>_s< \rho >_{env}< \sigma >_{fs}$

$< a_1 = a_2 \mapsto S>_s< \rho >_{env}< \sigma >_{fs} \rightarrow < a_2 \mapsto a_{1=} \; \square \mapsto S>_s< \rho >_{env}< \sigma >_{fs}$

$< if \; a \; then \; a_t \; else \; a_f \mapsto S>_s< \rho >_{env}< \sigma >_{fs} \rightarrow < a \mapsto if \; \square \; then \; a_t \; else \; a_f \mapsto S>_s< \rho >_{env}< \sigma >_{fs}$

$< if \; a \; then \; a_t \; else \; a_f \mapsto S>_s< \rho >_{env}< \sigma >_{fs} \rightarrow < a_t \mapsto S>_s< \rho >_{env}< \sigma >_{fs}$ $if \; a=1$

$< if \; a \; then \; a_t \; else \; a_f \mapsto S>_s< \rho >_{env}< \sigma >_{fs} \rightarrow < a_f \mapsto S>_s< \rho >_{env}< \sigma >_{fs}$ $if \; a=0$

$<while \; a \; do \; a_t \mapsto S>_s <\rho>_{env}<\sigma>_{fs} \rightarrow < if \; a \; then \; a_t; while \; a \; do \; a_t \; else \; skip \mapsto S>_s <\rho>_{env}<\sigma>_{fs}$

$<while \; a \; do \; a_t \mapsto S>_s <\rho>_{env}<\sigma>_{fs} \rightarrow < if \; a \; then \; \{a_t; while \; a \; do \; a_t \} \; else \; skip \mapsto S>_s <\rho>_{env}<\sigma>_{fs}$

$<open(f) \mapsto S>_s <\rho>_{env}<\sigma>_{fs} \rightarrow < S>_s <\rho>_{env}<\sigma>_{fs}$ $if \; \sigma(f)=c$

$<close(f) \mapsto S>_s <\rho>_{env}<\sigma>_{fs} \rightarrow < S>_s <\rho>_{env}<\sigma>_{fs}$ $if \; \sigma(f)=o$

$<(x,p)=read(f) \mapsto S>_s <\rho>_{env}<\sigma>_{fs} \rightarrow < S>_s <\rho[p\rightarrow n, x\rightarrow \varphi(f,n)]>_{env}<\sigma>_{fs}$ $if \; \sigma(f)=o$

$<S_1;S_2 \mapsto S>_s <\rho>_{env}<\sigma>_{fs} \rightarrow < S_1 \mapsto S_2 \mapsto S>_s <\rho>_{env}<\sigma>_{fs}$

$<fork\{S_1,S_2,\ldots,S_n\} \mapsto S>_s < \rho >_{env}< \sigma >_{fs} \rightarrow <a_{\theta(1)} ; a_{\theta(1)} ;\ldots; a_{\theta(n)} \mapsto S>_S <\rho>_{env} <\sigma>_{fs}$

$<forkfor\{S \} \mapsto S>_s < \rho >_{env}< \sigma >_{fs} \rightarrow <fork\{S,S,\ldots,S\} \mapsto S>_S <\rho>_{env} <\sigma>_{fs}$

$<forkif\{(S_1,a_1), (S_2,a_2),\ldots, (S_n,a_n)\} \mapsto S>_s < \rho >_{env}< \sigma >_{fs} \rightarrow <fork\{if \; a_1 \; then \; S_1 \; else \; skip, if \; a_2 \; then \; S_2 \; else \; skip ;\ldots; if \; a_n \; then \; S_n \; else \; skip \} \mapsto S>_S <\rho>_{env} <\sigma>_{fs}$

Fig. 1 $while_f$ complete semantics.

After explaining its structure, now we are ready to formally define the language $while_f$, as:

**Definition 3**. (*$while_f$, definition*) The language $while_f$ equals ($\sum_{whilef}$, $E_{whilef}$, $R_{whilef}$) (Figure 1). We say in $while_f$ that τ can be written to τ' if $while_f \vdash \tau \rightarrow^* \tau'$.

Here $\sum_{whilef}$ contains both the syntax of $while_f$ and ∑. Also both *E* and desugaring equations are in $E_{whilef}$. Since $\sum_{whilef}$ can be considered the algebraic specification of the language, the rewrite logic semantics of $while_f$ can be known as $while_f$.

## 3. File Safety

In this section we will define the file safety and termination of programs constructed in $while_f$. As we know the termination is un-decidable property in general, hence the file safety is also un-decidable. A program will be file safe if it reach a normal configuration under any possible executions. The choice of pointer values will play the major rule here since the read rule produces the un-determinism. This section will present the fact that the file safety is an un-decidable property even for terminating programs.

**Definition 4**.

For any computation *S* in $while_f$, we say that *S* terminates if and only if *[S]* is a terminating configuration in $while_f$, and we say that *S* is file safe if and only if any normal form of *S* is a final configuration in $while_f$.

We mentioned that file safety is not decidable even for terminating programs. One way to control the file safety is to add restriction on the *read* rule by letting the pointer value be one of its parameters. What we will do in the next section is to decide and tell where to read at a specific place. We may also treat it as in [21]. Now we state and proof the un-decidability of file safety even for terminating programs.

**Proposition 1:** File safety of terminating programs in $while_f$ is un-decidable property.

*Proof:*

Towards a contradiction we assume that the file safety is a decidable property. Since our language is Turing complete we can encode any decidable property $\mu(n)$ where $n \in Z$.

Let $y=n$; $PGM_\mu$ be a terminating and file safe program, this program writes a variable *out* such that $\mu(n)$ holds if and only if *out* = 1 in the environment and otherwise if *out* =0.

Since the pointer returned by reading from a file is nondeterministic in the following

*open(f);*

*forkfor{ (x,p) = read (f)}*

*close (f);*

We can use it to choose a random value for *y*.

We let

$PGM'_\mu \equiv$

*open (f)*

*forkfor{(x,y)= read(f)}*

*close (f)*; $PGM_\mu$;

$PGM'_\mu$ terminates for any *n* returned by *read*(f) and the loop always terminates; for file safety, $PGM'_\mu$ will be file safe if and only if *out*=1 when $PGM_\mu$ terminates which happens if and only if $\mu(n)$ holds for every $n \in Z$ [17].

To solve this problem we will control the pointer value and at the same time ignore the end of file error for simplicity. What we do to control the un-determinism that results from *read* rule is to take the pointer value as a parameter to read from a file. So instead of giving the file name to read from, and then get the value read and the pointer value, we will give the file name and the location to read from and get only the value read.

File safety also includes not opening a file that is already opened, not to close a file that is already closed, and reading from a closed file. All the previous cases are controlled with the side condition in the semantic rules of the language. That's why we use the file status entry.

## 4. Strong File Safety

This section presents the semantics of $SafeWhile_f$ language. The semantics will deal with the problem mentioned above. Also we will introduce the concept of *Strong file safety*. A program is strongly file safe if it reaches a final configuration in the language $SafeWhile_f$. The semantics changes the *read* rule and keeps all the other rules unchanged as Figure 2 shows. Under this modification the safety issue becomes decidable for terminating programs of the language $SafeWhile_f$. Those programs are strongly terminating. A program is strongly terminating if it terminates in $SafeWhile_f$. When a program is file safe according to the $SafeWhile_f$ rules we say it is *strongly file safe*.

$$<(x)=read(f,n) \mapsto S>_s <\rho>_{env}<\sigma>_{fs} \rightarrow < S>_s$$
$$<\rho[x \rightarrow \varphi(f,n)]>_{env}<\sigma>_{fs} \text{ if } \sigma(f)=o, \text{ Where } n \in N$$

Fig. 2 The semantics of $SafeWhile_f$.

All the other rules are the same as in $while_f$. The rule in Figure 2 is a *read* rule with the pointer value being given with the file name. First the rule has the side condition that the file status must be *o*. This means that the file must be opened before reading from it. If the file is closed then the rule will not be applied; the side condition avoids the issue of reading from a closed file which will cause an error. We now have to tell where to read from. So when reading we have to say we want to read from file *f* at location *n*. The pointer value is determined which will eliminate the un-determinism. This is the difference between $SafeWhile_f$ and $while_f$.

**Proposition 2:**
The rule *read* in $SafeWhile_f$ is deterministic.
*Proof*:
Since the pointer value is now considered as a parameter that should be giving along with the file name, the non-determinism is eliminated.

Now we can define the file safety and the termination according to $SafeWhile_f$.

**Definition 5:**
For any computation *S* in $SafeWhile_f$, we say that *S strongly terminates* if and only if *[S]* is a terminating configuration in $SafeWhile_f$, and we say that *S* is *strongly file safe* if and only if any normal form of *[S]* is a final configuration in $SafeWhile_f$.

**Proposition 3**:
Let $c \in S$ be a program, then:

1. If *c* is terminating then *c* is strongly terminating;
2. If *c* is strongly file safe then *c* is file safe;
3. If *c* is strongly file safe then *c* is terminating if and only if *c* is strongly terminating.

*Proof:*

1. Let c be a terminating program, suppose that $(\tau_i)_{i\geq 0}$ is a sequence of configurations such that $c=\tau_0$. Let $while_f \vdash \tau_i \xrightarrow{\gamma_{i+1}} \tau_{i+1}$, $i\geq 0$. Also suppose that $p_i$ is the sequence of pointers generated by $(\gamma_i)$ which is instance of *read* rules. We define a function $T: \Gamma_{whilef} \rightarrow \Gamma_{StrongWhilef}$ such that it maps every configuration to itself except for *read* rule. In case of *read* rule it eliminates all the pointers from the environment generated by the rule. $T(\tau)= <(x)=read(f,np) \mapsto S>_s <\rho/\{p_i\}>_{env}<\sigma>_{fs}$ if $\tau=<(x,p)=read(f) \mapsto S>_s <\rho>_{env}<\sigma>_{fs}$ and $T(\tau)=\tau$ otherwise.
   Now we will show that $T(\tau_i)$, $i\geq 0$ is $SafeWhile_f$ configuration, and $SafeWhile_f \vdash T(\tau_i) \xrightarrow{T(\gamma)_{i+1}} T(\tau_{i+1})$ ($T(\gamma_i)$ is an instance of $SafeWhile_f$ rule). Since $T(\tau_0)= \tau_0$, $\tau_0$ is a $SafeWhile_f$ configuration since the environment contains no pointer values. Suppose that $T(\tau_i)$ is a $SafeWhile_f$ configuration, and $while_f \vdash \tau_i \xrightarrow{\gamma_{i+1}} \tau_{i+1}$. We have two cases:
   Case 1: $\gamma_{i+1}$ is not an instance of *read* rule; In this case $\gamma_{i+1}$ is the same rule as in $SafeWhile_f$.
   Case 2: $\gamma_{i+1}$ is an instance of *read* rule; Then $T(\gamma_{i+1})$ is an instance of *read* rule in $SafeWhile_f$ by the definition of *T*.

2. Let $\tau_0$ be the initial state, and let $SafeWhile_f \vdash \tau_i \xrightarrow{\gamma_{i+1}} \tau_{i+1}$, $i\geq 0$, if $\tau_i$ is finite sequence of length *n* then, $\tau_n$ is a final configuration. Now we need to show that $while_f \vdash T'(\tau_i) \xrightarrow{T'(\gamma)_{i+1}} T'(\tau_{i+1})$ is either finite or it terminates at final configuration, such that $T'(\tau_0)= \tau_0$. We define *T'* as follows: $T'(\tau)= <(x,p)=read(f) \mapsto S>_s <\rho \cup \{p\rightarrow n\}>_{env}<\sigma>_{fs}$ if $\tau=<(x)=read(f,np) \mapsto S>_s <\rho>_{env}<\sigma>_{fs}$ and $T'(\tau)=\tau$ otherwise. Where $\{p_i\}$ is fresh variables. Now by the same way we can show that $while_f \vdash T'(\tau_i) \xrightarrow{T'(\gamma)_{i+1}} T'(\tau_{i+1})$ will be infinite or reach a final state from *T'* definition.

3. For the only if direction it follows from 1; for the if direction assume that *c* is strongly file safe and strongly terminating, then from 2 , *c* is file safe and by the same way it is terminating.

Finally we can conclude that the file safety is decidable only if the program is strongly terminating.
**Theorem:**
File safety of terminating programs is a decidable property for the language $SafeWhile_f$.
*Proof:*
If a program terminates in $SafeWhile_f$ then it has a finite number of derivation. Hence we can decide whether the last term is final or not.

The file safety is now a decidable property under the proposed condition. This special case enables us to check the file safety in the run time, for programs of $SafeWhile_f$.

## 5. Conclusion

File safety is un-decidable in general even for terminating programs of the language $while_f$. But when we tide up the pointer values the read command, it become decidable. The language $while_f$ is undeterministic because of the *read* command. The Read command return the pointer value along with the read value. We modified the language to get the language $SafeWhile_f$. The change is related to the syntax of the read command. The pointer value becomes required to read from a file. So to read from a file you need the file name and the location to read from. This makes file safety a decidable property for terminating programs in $SafeWhile_f$. So strong file safety of strongly terminating programs is decidable.

## Related Work

Work in the fields of files and operational semantics, are related to our work. The work in [11] checks memory safety in a dynamic way defining a semantics of language and changing it to make the problem decidable. The work in [21] repairs file problems with defining two type systems; one to admit the error and the other to repair it. In [2] the authors deal with the problem of memory safety via runtime systems via maintaining the soundness using probabilistic methods (probabilistic memory safety). The studied heap in this last paper has an infinite size. This last paper also introduces an algorithm that guarantees memory safety. The problem of memory safety is discussed also in [15] where the memory leaks are detected via software testing. The method in [15] depends on inserting checking points in the code produced by the compiler. These points check every memory access and detect the access errors. The work in [15] also tracks memory usage.

In the work [12] both type inference and run-time checking are combined to achieve type safety of programs via separating pointers according to their usage (static usage, run time usage). An attempt to automatically correct memory errors without the programmer interaction is made in [14] via deriving a runtime patches to fix the memory errors. These patches help in fixing bugs by merging the patches that come from multiple users. The language defined in [22] has a very similar structure to the one defined in [21]. Both used re-write logic as we do in the current paper.

In [9,8], approaches to correct programs via eliminate dead code [9,8] are introduced. An enrichment of the type system of pointers is introduced in [8] for live stack-heap analysis. The same author also in [6] works on the problem of memory safety and pointer analysis. In [6] the author introduces a flow sensitive type system for pointer analysis of multithreaded programs. The work in [7] presents a new flow sensitive technique for probabilistic pointer analysis in the form of type systems.

The work presented in [13] recovers a faulty process with a separation of data recovery enabling micro-rebooting. This technique recovers faulty application component without disturbing the rest of the application. This method is fast and cheap. Demsky and Dash introduce a language for robust software systems in [3]. This language has two levels. The first is a high level organization specification component, and the second is a low level operational specification. The high level is used to detect the errors, recover them, and to reason how to continue the execution safely. Model-based data structure repair is a technique introduced in [4], which enable a program to continue its execution in the face of data structure errors. The authors present an algorithm that use goal directed reasoning to translate model repair to data structure repair.

Denney and Fischer developed a frame work in [5] to allow proving the safety of a program statically and guarantee a dynamic safety. The violation asserted is used in [1] to repair the state of the program and to let it proceed. The authors present an algorithm to repair data structure with the structure that violates an assertion being given. The paper [16] describes also a method for software recovery from errors associated with a mechanism to preserve the information at a level of overhead, which is believed to be helpful. The concept of safe partial availability and safe partial anticipability, are the basic concepts for the algorithm introduced in [24]. The algorithm is based on flow graph and is better than the concept of safe partial availability. An algorithm also based on the concept of availability anticipability is introduced in [25]. This algorithm is designed for classical PRE to solve the problem of maximum flow. Saabas in [23] presents a program proof to be checked instead of checking the program code. Also the program carrying code [23] is used in two things; to match Hoare logic against low-level languages given a compositional semantics and to provide proof compilation.

## References

[1] B. Elkarablieh, I. Garcia, Y. L. Suen, and S. Khurshid.Assertion-based repair of complex datastructures. In Proc.of 22nd IEEE/ACM Int. Conf. on Automated Software Engineering,pp. 64–73. ACM Press, 2007.

[2] E. D. Berger and B. G. Zorn. Diehard: probabilistic memory safety for unsafe languages. In PLDI ’06: Proceedings of the 2006 ACM SIGPLAN conference on Programming language design and implementation, pages 158–168, New York, NY, USA, 2006. ACM.

[3] B. Demsky and A. Dash. Bristlecone: A language for robustsoftware systems. In J. Vitek, ed., Proc. of 22nd Europ. Conf. on Object-Oriented Program., ECOOP 2008, Lect. Notes in Comput. Sci., v. 5142, pp. 490–515. Springer, 2008.

[4] B. Demsky and M. C. Rinard. Goal-directed reasoning forspecification-based data structure repair. IEEE Trans. On Softw. Engin., 32(12):931–951, 2006.
[5] E. Denney and B. Fischer. Correctness of source-level safety policies. In K. Araki, S. Gnesi, and D. Mandrioli, eds., Proc.of 2003 Symp. of Formal Methods Europe, FME 2003, Lect. Notes in Comput. Sci., v. 2805, pp. 894–913. Springer, 2003.
[6] Mohamed A. El-Zawawy. (2011). Flow sensitive-insensitive pointer analysis based memory safety for multithreaded programs. In Beniamino Murgante, Osvaldo Gervasi, Andrés Iglesias, David Taniar, and Bernady O. Apduhan, editors, ICCSA (5), volume 6786 of Lecture Notes in Computer Science, pages 355-369. Springer.
[7] Mohamed A. El-Zawawy. (December 2011). Probabilistic pointer analysis for multithreaded programs. ScienceAsia, 37(4).
[8] Mohamed A. El-Zawawy. (January 2012). Dead code elimination based pointer analysis for multithreaded programs. Journal of the Egyptian Mathematical Society. doi:10.1016/j.joems.2011.12.011.
[9] Mohamed A. El-Zawawy. (March 2011). Program optimization based pointer analysis and live stack-heap analysis. International Journal of Computer Science Issues, 8(2).
[10] Mohamed A. El-Zawawy and Hamada A. Nayel. (October 2011). Partial redundancy elimination for multi-threaded programs. IJCSNS International Journal of Computer Science and Network Security, 11(10).
[11] Bernd Fischer, Ando Saab, and Tarmo Uustalu. Program Repair as Sound Optimization of Broken Programs. Springer,Verlag Berlin, Heidelberg, 2009.
[12] G. C. Necula, S. McPeak, and W. Weimer. CCured: type-safe retrofitting of legacy code. In POPL ’02: Proceedings of the 29th ACM SIGPLANSIGACT symposium on Principles of programming languages, pages 128–139, New York, NY, USA, 2002. ACM.
[13] G. Candea, S. Kawamoto, Y. Fujiki, G. Friedman, and A. Fox. Microreboot—a technique for cheap recovery. In Proc. of 6th Symp. on Operating System Design and implementation,OSDI 2004, pp. 31–44. Usenix Assoc., 2004.
[14] G. Novark, E. D. Berger, and B. G. Zorn. Exterminator: Automatically correcting memory errors with high probability. Commun. ACM, 51(12):87– 95, 2008.
[15] R. Hastings and B. Joyce. Purify: Fast detection of memory leaks and access errors. In Proceedings of the Winter USENIX Conference.
[16] J. J. Horning, H. C. Lauer, P. M. Melliar-Smith, and B. Randell. A program structure for error detection and recovery. In Proc. of Symp. on Operating Systems, Lect. Notes in Comput. Sci., v. 16, pp. 171–187. Springer, 1974.
[17] H. Rogers Jr. Theory of Recursive Functions and Effective Computability. MIT press, Cambridge, MA, 1987.
[18] J. Meseguer. Conditioned rewriting logic as a united model of concurrency. Theor. Comput. Sci., 96(1):73–155, 1992.
[19] J. Meseguer and G. Ro¸su. The rewriting logic semantics project. Theor. Computer Science, 373(3):213–237, 2007.
[20] T. F. ¸Serb˘anu¸t˘a, G. Ro¸su, and J. Meseguer. A rewriting logic approach to operational semantics. Inf. and Comp., 2009. to appear; http://dx.doi.org/10.1016/j.ic.2008.03.026.
[21] Grigore Rosu, Wolfram Schulte, and Traian Florin Serbanuta. Runtime verification of C memory safety. Springer,Verlag Berlin, Heidelberg, 2009.
[22] G. Ro¸su. K: A Rewriting-Based Framework for Computations – Preliminary version. Technical Report UIUCDCS-R-2007-2926, University of Illinois, 2007.
[23] A. Saabas. Logics for low-level code and proof-preserving program transformations (PhD thesis), Thesis on Inform. and Syst. Engin. C43. Tallinn Univ. of Techn., 2008.
[24] V. K. Paleri, Y. N. Srikant, and P. Shankar. Partial redundancy elimination: a simple, pragmatic, and provably correct algorithm. Sci. of Comput. Program., 48(1):1–20, 2003.
[25] J. Xue and J. Knoop. A fresh look at PRE as a maximum flow problem. In A. Mycroft and A. Zeller, eds., Proc. Of 15th Int. Conf. on Compiler Construction, CC 2006, Lect. Notes in Comput. Sci., v. 3923, pp. 139–154. Springer, 2006.

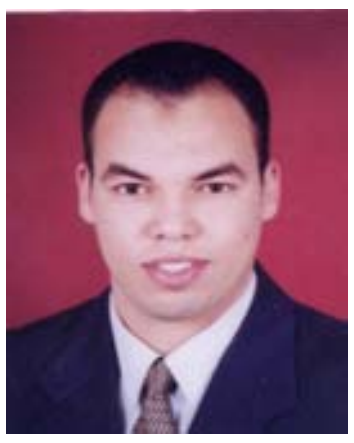

**Dr. Mohamed A. El-Zawawy** received: PhD in Computer Science from the University of Birmingham in 2007, M.Sc. in Computational Sciences in 2002 from Cairo University, and a BSc. in Computer Science in 1999 from Cairo University. Dr El-Zawawy is an assistant professor of Computer Science at Faculty of Science, Cairo University Since 2007. Currently, Dr. El-Zawawy is on a sabbatical from Cairo University to College of Computer and Information Sciences, Al-Imam M. I.-S. I. University, Riyadh, Kingdom of Saudi Arabia. During the year 2009, Dr. El-Zawawy held the position of an extra-ordinary senior research at the Institute of Cybernetics, Tallinn University of Technology, Estonia. Dr. El-Zawawy worked as a teaching assistant at Cairo University from 1999 to 2003 and latter at Birmingham University from 2003 to 2007. Dr. El-Zawawy is interested in static analysis, shape analysis, type systems, and semantics of programming languages.